\documentclass[sigplan,10pt]{acmart}
\usepackage{xcolor}
\usepackage{tcolorbox}

\newif\ifshowcomments
\showcommentstrue

\renewcommand\footnotetextcopyrightpermission[1]{}
\AtBeginDocument{%
  }

\usepackage{xspace}
\usepackage{xcolor}
\usepackage{microtype}
\usepackage{array}
\usepackage{float}
\usepackage{tikz}
\usetikzlibrary{arrows.meta,positioning,decorations.pathreplacing}
\usepackage{listings}
\usepackage{lstlinebgrd}

\definecolor{lstkw}{HTML}{0033B3}
\definecolor{lstcmt}{HTML}{6A737D}
\definecolor{lststr}{HTML}{067D17}
\definecolor{lstnum}{HTML}{AAAAAA}
\definecolor{lsthl}{HTML}{FFF3C4}
\definecolor{lstgood}{HTML}{1B7F2E}
\definecolor{lstbad}{HTML}{C62828}

\lstdefinestyle{bpfix}{%
  basicstyle=\ttfamily\fontsize{6}{7}\selectfont,
  numbers=left, numberstyle=\fontsize{6}{7}\selectfont\color{lstnum},
  keywordstyle=\color{lstkw}\bfseries,
  commentstyle=\color{lstcmt}\itshape,
  stringstyle=\color{lststr}, showstringspaces=false,
  frame=single, framesep=2pt, rulecolor=\color{black!25},
  columns=fixed, basewidth=0.5em,
  breaklines=true, breakatwhitespace=true,
  numbersep=4pt,
  xleftmargin=1.1em, xrightmargin=0.4em, aboveskip=3pt, belowskip=3pt,
}

\definecolor{lstlogbg}{HTML}{EDEFF2}
\lstdefinestyle{bpflog}{%
  language=bash,
  basicstyle=\ttfamily\fontsize{6}{7}\selectfont,
  backgroundcolor=\color{lstlogbg},
  numbers=none,
  keywordstyle=\color{lstkw}\bfseries,
  frame=single, framesep=2pt, rulecolor=\color{black!25},
  columns=fixed, basewidth=0.5em,
  breaklines=true, breakatwhitespace=true,
  xleftmargin=1.1em, xrightmargin=0.4em, aboveskip=1pt, belowskip=3pt,
}

\definecolor{lstdiagbg}{HTML}{EAF4EC}
\lstdefinestyle{bpfdiag}{%
  basicstyle=\ttfamily\fontsize{6}{7}\selectfont,
  backgroundcolor=\color{lstdiagbg},
  numbers=none,
  frame=single, framesep=2pt, rulecolor=\color{black!25},
  columns=fixed, basewidth=0.5em,
  breaklines=true, breakatwhitespace=true,
  xleftmargin=1.1em, xrightmargin=0.4em, aboveskip=1pt, belowskip=3pt,
} 
\begin{document}

\title{Characterizing and Bridging the Diagnostic Gap in eBPF Verifier Rejections}
\acmCodeLink{https://github.com/eunomia-bpf/bpfix}

\author{Yusheng Zheng$^{*1,2}$, Zhengjie Ji$^{*3}$, Weichen Tao$^{*4}$, Xiangyu Gao$^{5}$, Jianchang Su$^{6}$, Wei Zhang$^{6}$, Andi Quinn$^{1}$, Dan Williams$^{3}$}
\affiliation{%
  \institution{$^{1}$UC Santa Cruz \quad $^{2}$eunomia-bpf \quad $^{3}$Virginia Tech \quad $^{4}$Telecom Paris \quad $^{5}$University of Washington \quad $^{6}$University of Connecticut}
  \country{}}
\renewcommand{\authorsaddresses}{}

\newcommand{\zj} [1]{\textcolor{blue}{{}#1}}
\newcommand{\wt}[1]{%
\ifshowcomments
{\color{green!40!gray}[weichen: #1]}
\fi
}
\newcommand{\note} [1]{\textcolor{red}{#1}}
\newcommand{\todo} [1]{\textcolor{red}{#1}}
\newcommand{\revise} [1]{\textcolor{orange}{#1}}
\newcommand{\revised} [1]{\textcolor{green!50!gray}{#1}}

\newcommand{\sys}{\texttt{bpfix}\xspace}
\newcommand{\empiricalset}{\texttt{bpfix-empirical}\xspace}
\newcommand{\bench}{\texttt{bpfix-bench}\xspace}

\newcommand{\citehere}[1][?]{%
  \textcolor{red}{[#1]}%
}
\begin{abstract}
eBPF lets developers run custom programs inside the Linux kernel, where a verifier proves each program safe.
However, when the verifier rejects a program, the unclear error makes repair challenging: the error reports where verification stopped, not where the program lost the proof the verifier required.
To quantify this gap, we conduct an empirical study of 235 reproduced rejections, showing that 47\% of rejections return only \texttt{EINVAL}, one error string maps to as many as nine distinct root causes, and 10 of the 12 root causes are eBPF-specific. Repair thus requires both domain knowledge and locating where the proof was lost, yet existing tools only help developers read the error.
We present \sys, which reconstructs where the required proof was established and where it was lost from the verifier log, and prints a Rust-like diagnostic.
To evaluate \sys and the ability of LLMs to help repair, we construct a benchmark of 75 LLM repair tasks. Current models achieve 0--37\% one-shot success with the raw log, and replacing the log with the \sys localization improves repair by 11--21pp, suggesting that locating where the proof was lost is key to guiding repair.
bpfix is available at \url{https://github.com/eunomia-bpf/bpfix}.
\end{abstract}

\settopmatter{printfolios=true,printacmref=false}
\maketitle
\begingroup
\renewcommand\thefootnote{}\footnotetext{$^{*}$Equal contribution.}%
\addtocounter{footnote}{-1}%
\endgroup
\pagestyle{plain}

\begin{figure}[t]
  \centering
  \includegraphics[width=0.96\columnwidth]{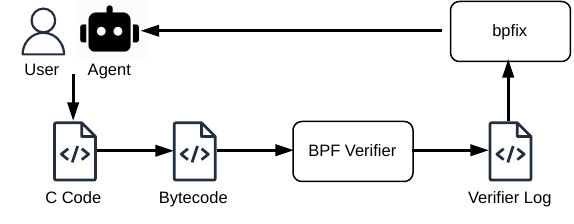}
  \vspace{0.5pt}

  \noindent\textbf{\footnotesize (a) C Code}
  \begin{lstlisting}[style=bpfix,language=C,firstnumber=263]
if (ipv4_hdr)
    udph = (void *)ipv4_hdr + sizeof(*ipv4_hdr);
else
    udph = (void *)ipv6_hdr + sizeof(*ipv6_hdr);
if (udph + sizeof(struct udphdr) > data_end)  // bounds-checked
    return 1;

dst_port = __constant_ntohs(((struct udphdr *)udph)->dest); // rejected
  \end{lstlisting}

  \noindent\textbf{\footnotesize (b) Verifier Log}
  \begin{lstlisting}[style=bpflog]
from 31 to 34: ... R5_w=40 ...
; if (udph + sizeof(struct udphdr) > data_end) @ prog.c:267
35: (07) r3 += 8                      ; R3=48
36: (2d) if r3 > r2 goto pc+4         ; R2=pkt_end() R3=48
; dst_port = ...((struct udphdr *)udph)->dest; @ prog.c:270
37: (69) r2 = *(u16 *)(r5 +2)
R5 invalid mem access 'scalar'
  \end{lstlisting}

  \noindent\textbf{\footnotesize (c) \sys output}
  \begin{lstlisting}[style=bpfdiag]
error[BPFOCUS-E006]: verifier-visible compiler lowering hides the required proof
  = class: lowering_artifact
  = confidence: medium
  = next action: provenance
  --> prog.c:270
   |
263 | if (ipv4_hdr)
    | ------------- nearby source context for pointer provenance
267 | if (udph + sizeof(struct udphdr) > data_end)
    | -------------------------------------------- verifier state changes from pkt to scalar before the rejected access
270 | dst_port = __constant_ntohs(((struct udphdr *)udph)->dest);
    | ^^^^^^^^^^^^^^^^^^^^^^^^^^^^^^^^^^^^^^^^^^^^^^^^^^^^^^^^^^^ rejected here: verifier sees a scalar where a pointer is required
   |
   = verifier[229]: R5 invalid mem access 'scalar'
   = required proof: preserve a verifier-recognized pointer type at the operation that requires a pointer
help: Reacquire a verifier-tracked pointer before the rejected dereference.
help: Use the packet pointer that received the data_end proof, or rederive and recheck it before the load.
  \end{lstlisting}
  \caption{\sys on a real rejection (Stack Overflow 53136145). The verifier returns one register-level line; from the same log, \sys reconstructs the discarded proof and localizes the rejection to the point where the required proof was lost, mapping each span back to source.}
  \label{fig:overview}
\end{figure}

\section{Introduction}
\label{sec:introduction}

eBPF lets developers run custom programs inside the Linux kernel for networking, security, and observability~\cite{gbadamosi2024ebpf, cilium}.
To load a program, a developer writes it in C or Rust, compiles it to eBPF bytecode, and submits it to a verifier. The verifier proves the program safe by tracking abstract values along every execution path~\cite{gershuni2019simple, linuxkernel-bpf-verifier}.
However, the unclear verifier rejection makes repair challenging.
Figure~\ref{fig:overview} illustrates this diagnostic gap: a packet pointer that the program bounds-checks is rejected at a later load with \texttt{R5 invalid mem access `scalar'}, on a line the developer has no reason to suspect.
Without a way to locate where the proof was lost, developers must rely on repeated trial and error~\cite{deokar2024empirical}.
To quantify this gap, we conduct an empirical study of 235 rejections reproduced from Stack Overflow, GitHub issues, fix commits, and kernel selftests.
We find that 191 are program bugs spanning 12 root causes, while the other 44 reject correct source due to compiler optimizations that obscure pointer types, environment misconfiguration, or verifier limitations.
Of the 12 root causes, 10 are eBPF-specific (e.g., packet bounds, dynptr lifetime), so repair requires domain knowledge beyond general programming.
The terminal error provides little help: 47\% of rejections return \texttt{EINVAL}, and one error string maps to as many as nine distinct root causes.
Existing tools do not close this gap.
PrettyVerifier~\cite{rizza2025design} and the BPF Verifier Visualizer~\cite{bpfvv} help developers read the error with regular expressions or visualizations, but neither locates where the proof was lost.
Error Explainers in other domains like Rust walk a structure the checker builds, such as an AST or typing constraints~\cite{pavlinovic2014finding, zhang2014toward}, but the eBPF verifier exposes no such artifact.
To improve the diagnostic process, we introduce \sys, which reconstructs where the required proof was established and where it was lost from the verifier log, and prints a Rust-like diagnostic (Figure~\ref{fig:overview}).
To measure the ability of LLMs to fix verifier rejections, we construct \bench, a benchmark of 75 repair tasks.
With the raw verifier log, current LLMs achieve only 0--37\% one-shot repair across three models ranging from 3B to 27B parameters, showing that verifier rejections remain difficult even for capable models.
To see whether better diagnostics help automated repair, we replace the log with the \sys localization, which improves repair by 11--21pp, with the gain largest at the stages where a repair must restore the verifier-visible proof.
The result suggests that locating where the proof was lost, rather than only where verification stopped, is key to guiding repair.
This paper makes the following contributions:
\begin{itemize}
\item An empirical study of 235 reproduced rejections showing the terminal error reports where verification stopped, leaving the developer to find where the proof was lost~(\S\ref{sec:study}).
\item \sys, a technique that reconstructs where the required proof was established and lost from the verifier log~(\S\ref{sec:design}).
\item \bench, a benchmark of 75 repair tasks for evaluating LLM-based verifier repair~(\S\ref{sec:benchmark}).
\end{itemize}

\section{An Empirical Study of Verifier Rejections}
\label{sec:study}

A rejection confronts the developer with a low-level log, and this section studies what that log records about real rejections and where it falls short.
We describe how a developer reads a rejection (\S\ref{sec:study:verifier}), and then we assemble \empiricalset, a dataset of real rejections (\S\ref{sec:study:corpus}), characterize the rejections in it (\S\ref{sec:study:bugs}), and measure the log's diagnostic gap (\S\ref{sec:study:gap}).

\subsection{Background: The Verifier Rejection}
\label{sec:study:verifier}

Before the kernel runs an eBPF program, the verifier must prove the program safe.
The verifier proves safety one instruction at a time, building for each value the proof that it is used safely, such as a packet pointer staying within bounds, and exploring every path.
A program that fails a check is rejected, and the verifier returns a log of the abstract state after each instruction, ending in a one-line terminal error at the rejected instruction.

The terminal error names the rejection location, the operation where verification stopped, which the verifier reaches only after the program lost the proof at an earlier instruction.
To repair the program so it passes the verifier, the developer must learn which proof the verifier required and where the program failed to keep it, because a fix has to restore that proof.
The error names only the operation that needed the proof, so the developer reconstructs the required proof and its loss by hand from the trace.
\begin{figure}[t]
  \begin{lstlisting}[style=bpfix,language=C,
      linebackgroundcolor={\ifnum\value{lstnumber}=3 \color{lstgood!40}\fi}]
// correct source; -O0 lowers the field read into a scalar
static int add_one(int x) { return x + 1; }
int prog(struct xdp_md *ctx) { return add_one(ctx->rx_queue_index) & 1; }
  \end{lstlisting}
  \begin{lstlisting}[style=bpflog]
verifier: R2 invalid mem access 'scalar'
  \end{lstlisting}
  \caption{A compiler-lowering rejection: correct source that the verifier reports as \texttt{invalid mem access `scalar'}.}
  \label{fig:packet}
\end{figure}
 
\subsection{Reproducing Real Rejections in \empiricalset}
\label{sec:study:corpus}

To study rejections that developers actually hit, we assemble \empiricalset, a dataset of real verifier rejections reproduced under one fixed toolchain.
The dataset draws on 936 candidate reports from four sources: Stack Overflow questions, GitHub issues, GitHub fix commits, and kernel selftests.
We rebuild each candidate and load it into kernel 6.15.11 with clang 18 at log level 2, and 235 of the 936 reject under this build and form \empiricalset.
The rest are dropped because they do not reject under our toolchain, need a specific environment, or lack the source to rebuild.

Each \empiricalset case carries the faulty source and the developer's own fix, both from its report.
The source and the fix give the ground truth for the rejection and where its repair landed, which the next two subsections use.

\subsection{Characterizing the Rejections}
\label{sec:study:bugs}

We characterize each \empiricalset rejection by where its repair lands.
In 191 of the 235 rejections the repair changes the program's source.
The other 44 reject correct source and are repaired elsewhere: in the compiler (18), the environment (14), or the verifier (12).
For example, a context-field read that is provably safe is rejected because \texttt{-O0} lowers it into a scalar.
The repair is a compiler flag that leaves the source untouched (Figure~\ref{fig:packet}).
The rest of this subsection examines the 191 program bugs.

A program bug fails a safety check, but the failed check does not name the source mistake.
We therefore label each bug with its root cause, the source-level error the developer made.
The 191 bugs fall into 12 root causes (Table~\ref{tab:mechanisms}), most specific to eBPF, such as proving a packet bound or pairing a dynptr with its lifetime.

\begin{tcolorbox}[colback=blue!5!white,colframe=gray!75!black,left=1mm, right=1mm, top=0.5mm, bottom=0.5mm, arc=1mm]
\textbf{Takeaway \#1:} Most rejections (81\%) are program bugs, but 19\% reject correct source due to compiler, environment, or verifier issues. Of 12 root causes, 10 are eBPF-specific, so repair requires domain knowledge.
\end{tcolorbox}

\begin{table}[t]
  \centering
  \caption{The 191 developer bugs fall into 12 root-cause categories. A category is the specific source-level error the developer made; the \emph{Cases} column gives how many bugs each accounts for.}
  \label{tab:mechanisms}
  \small
  \begin{tabular}{@{}l r@{}}
    \toprule
    Root-cause category & Cases \\
    \midrule
    Unclamped scalar used as offset or length   & 24 \\
    Corrupted or stale dynptr object            & 23 \\
    Packet access without a bound on every path  & 22 \\
    Missing null check                           & 19 \\
    Pointer type or provenance mismatch          & 16 \\
    Unverified address dereferenced              & 16 \\
    Index exceeds object capacity                & 15 \\
    Context or contract misuse                   & 15 \\
    Unpaired resource reference                  & 15 \\
    Interrupt flag not restored in order         & 11 \\
    Probe signature mismatched with the ABI      & 9  \\
    Oversized or uninitialized stack buffer      & 6  \\
    \midrule
    Total                                        & 191 \\
    \bottomrule
  \end{tabular}
\end{table} 
\subsection{Measuring the Log's Localization Gap}
\label{sec:study:gap}

With the root causes labeled, we find the terminal error too coarse to name which one a rejection hit.
The errno it carries does little to narrow the cause, as \texttt{EINVAL} alone tags 47\% of all rejections.
The error string is just as coarse, because one string maps to many root causes.
To measure how often, we mask the registers and offsets in each string and group the 235 rejections by the normalized template.
This collapses 167 distinct strings into 82 templates.
Fifteen of the 82 templates each map to more than one root cause.
The most frequent alone covers nine root causes across 28 rejections (Table~\ref{tab:fanout}).
\begin{table}[t]
  \centering
  \caption{The four most frequent message templates, with the cases sharing each template and the distinct root-cause categories they span. A template normalizes registers and offsets to placeholders.}
  \label{tab:fanout}
  \small
  \begin{tabular}{@{}l r r@{}}
    \toprule
    Terminal message template & Cases & Categories \\
    \midrule
    \texttt{R\# invalid mem access `scalar'}   & 28 & 9 \\
    \texttt{invalid access to packet}          & 26 & 5 \\
    \texttt{invalid access to map value}       & 18 & 4 \\
    \texttt{R\# !read\_ok}                     & 13 & 4 \\
    \bottomrule
  \end{tabular}
\end{table} 
Because one error matches many root causes, the error alone cannot identify which proof the program lost, or where, the two facts a repair needs.
The trace, however, holds more than the terminal error.
Under \texttt{log\_level=2} the verifier prints its abstract state after every instruction, so the lost proof survives in the printed states, and \S\ref{sec:design} reconstructs the proof from those states to localize the loss.

\begin{tcolorbox}[colback=blue!5!white,colframe=gray!75!black,left=1mm, right=1mm, top=0.5mm, bottom=0.5mm, arc=1mm]
\textbf{Takeaway \#2:} The terminal error is too coarse to guide repair: 47\% of rejections return \texttt{EINVAL}, and one error string maps to as many as nine root causes.
\end{tcolorbox}

\section{\sys Design and Implementation}
\label{sec:design}

\begin{figure}[t]
    \centering
    \includegraphics[width=\columnwidth]{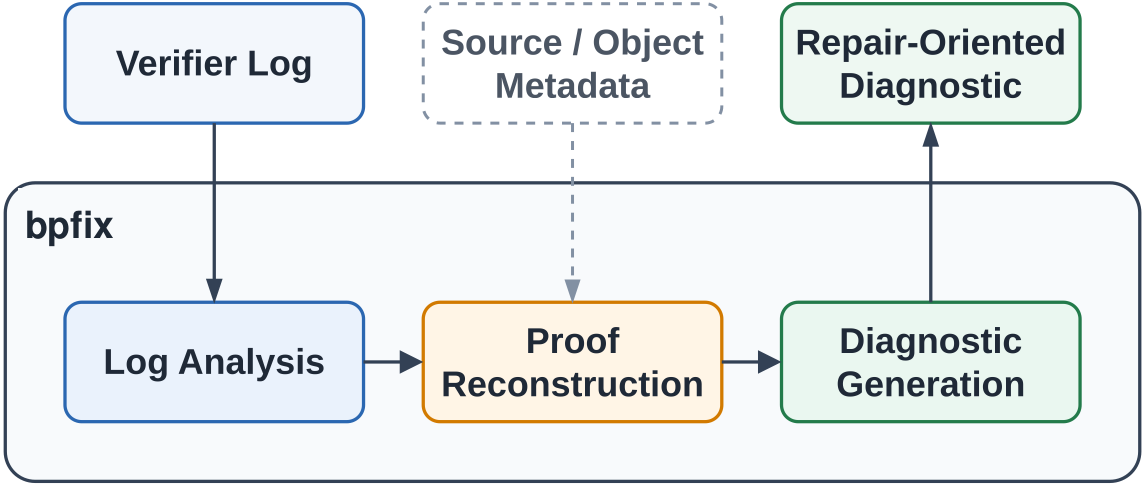}
    \caption{Overview of the \sys workflow. \sys analyzes verifier logs, reconstructs the missing proof, and generates a repair-oriented diagnostic.}
    \label{fig:design-overview}
\end{figure} %
The goal of \sys is to answer two questions the terminal error leaves open: which proof the verifier required, and where the program lost it.
\sys reconstructs the proof lifecycle from the existing verifier log by parsing the per-instruction abstract states and tracking when evidence for the required proof appears, persists, and is lost.
Figure~\ref{fig:design-overview} summarizes the three-stage workflow: log analysis normalizes evidence around the rejection, proof reconstruction identifies and tracks the required proof, and diagnostic generation turns the resulting evidence into guidance the developer can act on.

\subsection{Architecture and Data Flow}
\label{sec:design-architecture}

\sys is implemented in approximately 23k lines of Rust as a pipeline whose only required input is an existing verifier log; optional source or object metadata improves instruction-to-source mapping.
The Log Analysis stage extracts the terminal error and the per-instruction abstract states into a normalized evidence stream.
The Proof Reconstruction stage maps the terminal error to a \emph{proof family} (such as pointer provenance, packet bounds, or scalar range), identifies the required proof, and tracks evidence for that proof.
The Diagnostic Generation stage selects spans and derives guidance for re-establishing the proof.

The result is a Rust-style plain-text diagnostic with a stable error identifier, the required proof, relevant spans and evidence, a loss point when observable, and \texttt{help:} guidance.

\subsection{Reconstructing the Verifier-Visible Proof}
\label{sec:design-proof}

\sys first identifies which proof the verifier required.
The terminal error identifies the rejected operation and immediate symptom, while nearby abstract state supplies parameters such as the affected register, access offset and size, scalar range, pointer type, or reference identifier.
Together they identify the required proof at the \emph{rejection location}, the instruction where the verifier stopped.
For example, in Figure~\ref{fig:overview}, \texttt{R5 invalid mem access `scalar'} maps to a pointer-provenance proof.
From the rejected load and nearby state, \sys infers that \texttt{R5} must still carry verifier-recognized packet-pointer provenance when the rejected load reads it.

\sys then tracks evidence for that proof through the log: when the evidence appears (establishment), when it becomes incompatible (loss), and when the verifier requires it (rejection).
When both establishment and loss are observed, the transition between them becomes the \emph{loss point}.
In the running example, an earlier state represents \texttt{R5} as \texttt{pkt(off=34,r=42)}, whereas a later state before instruction~37 represents it as scalar value \texttt{40}.
The final load dereferences \texttt{R5}, so \sys reports the observed pointer-to-scalar transition as the loss point and instruction~37 as the rejection location.

\subsection{Evidence-Aware Diagnostic Synthesis}
\label{sec:design-synthesis}

Diagnostic synthesis adapts its guidance to the reconstructed proof lifecycle.
\sys localizes the evidence needed for repair: the rejected operation, the required proof at that operation, the loss point when the log exposes one, and the span and next action that guide the developer in re-establishing the proof.
When the proof was never established, \sys recommends introducing it, for example through a bounds check, null check, or reference acquisition.
When the proof is visible and later lost, it recommends preserving or re-deriving it near the rejected use.
Developers use the resulting guidance to choose and validate a repair through the normal compile-and-load workflow.
\section{Evaluation}
\label{sec:evaluation}

This section evaluates the \sys diagnostic with two case studies on \empiricalset, the 235 reproduced rejections of \S\ref{sec:study:corpus}.
The first shows how the diagnostic recovers the repair information the terminal error omits, and the second shows how it separates the responsible layer behind a shared terminal error.

\subsection{Recovering the Missing Repair Information}
\label{sec:eval-diagnostic-completeness}

We examine one rejection from the aya project to show how \sys recovers the repair information the terminal error omits.

\begin{figure}[t]
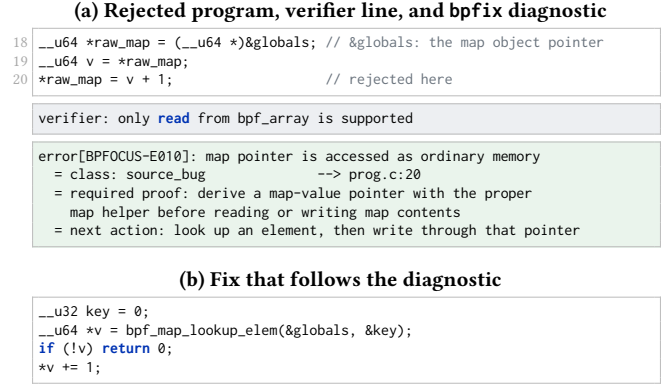

  \noindent\textbf{\footnotesize (a) Rejected program, verifier line, and \sys diagnostic}
  \begin{lstlisting}[style=bpfix,language=C,firstnumber=18]
__u64 *raw_map = (__u64 *)&globals; // &globals: the map object pointer
__u64 v = *raw_map;
*raw_map = v + 1;                   // rejected here
  \end{lstlisting}
  \begin{lstlisting}[style=bpflog]
verifier: only read from bpf_array is supported
  \end{lstlisting}
  \begin{lstlisting}[style=bpfdiag]
error[BPFOCUS-E010]: map pointer is accessed as ordinary memory
  = class: source_bug              --> prog.c:20
  = required proof: derive a map-value pointer with the proper
    map helper before reading or writing map contents
  = next action: look up an element, then write through that pointer
  \end{lstlisting}
  \vspace{2pt}
  \noindent\textbf{\footnotesize (b) Fix that follows the diagnostic}
  \begin{lstlisting}[style=bpfix,language=C,numbers=none]
__u32 key = 0;
__u64 *v = bpf_map_lookup_elem(&globals, &key);
if (!v) return 0;
*v += 1;
  \end{lstlisting}
  \caption{A real rejection from the aya project (issue 1002): (a) the rejected program, the verifier's terminal line, and the \sys diagnostic; (b) the developer's fix.}
  \label{fig:diag-example}
\end{figure}

As shown in Figure~\ref{fig:diag-example}, the program treats the map object \texttt{\&globals} as a map-value pointer, casting its address to \texttt{\_\_u64 *} and reading and writing through the result.
The verifier rejects the write with the terminal line \texttt{only read from bpf\_array is supported}.
The line is correct, but it gives only the symptom: the verifier sees a write through the map object where a map value is required.

From the same log, \sys reconstructs the lost proof.
The diagnostic states the proof the rejected write needs, a map-value pointer derived through a map helper before any access to map contents.
It then localizes the loss to the cast that produced a raw map-object pointer.
As shown in Figure~\ref{fig:diag-example}, the developer's fix follows that proof: it calls \texttt{bpf\_map\_lookup\_elem}, checks the returned pointer, and updates the value through the result.
The terminal line gives only the rejected operation, while the diagnostic recovers the lost proof and the point where the program lost it.
 
\subsection{Separating the Responsible Layer}
\label{sec:eval-layer-attribution}

A second diagnostic gap is that one terminal verifier message can call for repairs in different layers.
\sys reports the responsible layer as a best-effort attribution, labeling a rejection \texttt{source\_bug} for the source layer of \S\ref{sec:study} or \texttt{lowering\_artifact} for the compiler layer. We inspect whether \sys separates two rejections that share one terminal error but belong to different layers.

\begin{figure}[t]
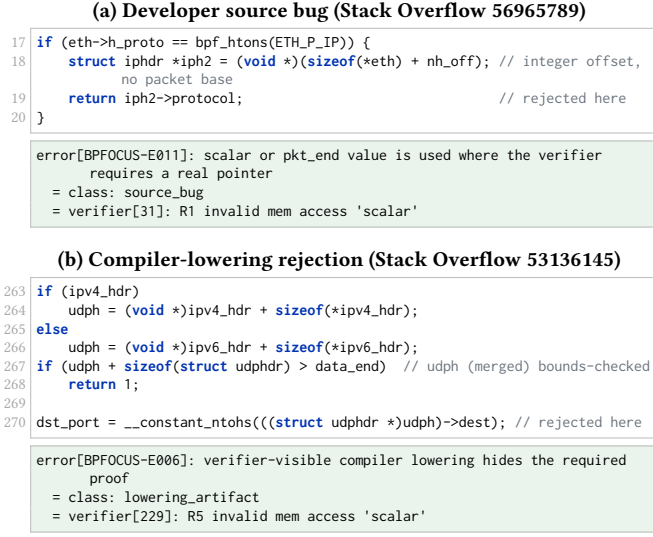

  \noindent\textbf{\footnotesize (a) Developer source bug (Stack Overflow 56965789)}
  \begin{lstlisting}[style=bpfix,language=C,firstnumber=17]
if (eth->h_proto == bpf_htons(ETH_P_IP)) {
    struct iphdr *iph2 = (void *)(sizeof(*eth) + nh_off); // integer offset, no packet base
    return iph2->protocol;                                // rejected here
}
  \end{lstlisting}
  \begin{lstlisting}[style=bpfdiag]
error[BPFOCUS-E011]: scalar or pkt_end value is used where the verifier requires a real pointer
  = class: source_bug
  = verifier[31]: R1 invalid mem access 'scalar'
  \end{lstlisting}
  \vspace{2pt}
  \noindent\textbf{\footnotesize (b) Compiler-lowering rejection (Stack Overflow 53136145)}
  \begin{lstlisting}[style=bpfix,language=C,firstnumber=263]
if (ipv4_hdr)
    udph = (void *)ipv4_hdr + sizeof(*ipv4_hdr);
else
    udph = (void *)ipv6_hdr + sizeof(*ipv6_hdr);
if (udph + sizeof(struct udphdr) > data_end)  // udph (merged) bounds-checked
    return 1;

dst_port = __constant_ntohs(((struct udphdr *)udph)->dest); // rejected here
  \end{lstlisting}
  \begin{lstlisting}[style=bpfdiag]
error[BPFOCUS-E006]: verifier-visible compiler lowering hides the required proof
  = class: lowering_artifact
  = verifier[229]: R5 invalid mem access 'scalar'
  \end{lstlisting}
  \caption{Two rejections that share the verifier message \texttt{invalid mem access `scalar'} but need fixes in different layers: a source bug (a) and a compiler-lowering artifact (b). Each diagnostic is abridged to its error id, class, and the verifier line it explains.}
  \label{fig:layer-attr}
\end{figure}

Figure~\ref{fig:layer-attr} shows two rejections that both end with \texttt{invalid mem access `scalar'}.
In Figure~\ref{fig:layer-attr}(a), the program constructs a packet pointer by casting an integer offset, \texttt{(void *)(sizeof(*eth) + nh\_off)}.
The resulting value has no verifier-tracked packet base, so the later packet-field load uses a scalar where the verifier requires a real pointer.
\sys reports \texttt{source\_bug}: the source never establishes the pointer provenance the rejected load requires.

In Figure~\ref{fig:layer-attr}(b), the source expresses the intended packet-bound proof, deriving \texttt{udph} from an IPv4 or IPv6 header and bounds-checking it against \texttt{data\_end} before the read.
The rejection arises after compiler lowering merges the branch and hides the verifier-recognized packet-pointer type before the load.
The upstream fix steers the compiler and leaves the source logic unchanged, so the layer label holds independently of \sys.
\sys reports the case as \texttt{lowering\_artifact}.
On these two rejections, \sys reports a different layer for each, separating a source bug from a lowering artifact behind one terminal error.
 
\section{\bench: A Benchmark for LLM Verifier Repair}
\label{sec:benchmark}

To measure the ability of LLMs to fix verifier rejections and evaluate whether \sys helps, we construct \bench, a benchmark of 75 source-level repair tasks.

\subsection{Benchmark Construction}
\label{sec:benchmark-construction}

\bench contains 75 repair tasks: 40 built around a required verifier proof and 35 minimized from open-source projects such as Cilium~\cite{cilium}, xdp-tools~\cite{xdptools}, and bpftime~\cite{zheng2025extending}.
Each task ships the buggy program with an executable test suite independent of \sys, which accepts a fix only if it loads through the kernel verifier and passes the case's functional and source-semantics checks.
Because the tests are independent of \sys, a fix earns credit only by passing the same checks a correct program must pass.

We score each task in two modes.
The one-shot mode tests the diagnostic on its own, and the retry mode allows one failure-informed retry to test whether the help persists.
Our primary model is Qwen3.6 27B, and we also run the hosted GLM 5.2 and the smaller Qwen2.5 3B at temperature zero.
We include the small 3B model to see whether the gain survives at low model capacity.
For each task, a model receives the buggy program with either the raw verifier log or the \sys diagnostic and writes a candidate fix for the test suite to judge.

\subsection{Results}
\label{sec:benchmark-results}

\begin{figure}[H]
  \centering
  \includegraphics[width=0.9\columnwidth]{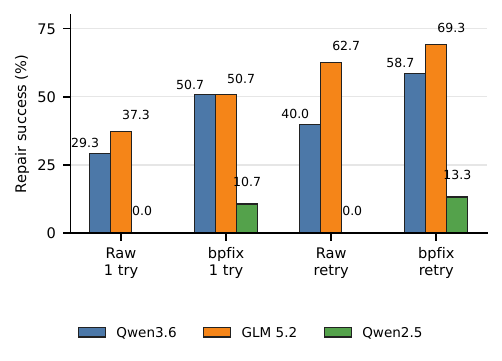}
  \caption{\bench repair success across three models. Bars are grouped by prompt mode and colored by model; retry bars use one failure-informed retry.}
  \label{fig:bpfix-bench}
\end{figure}

Replacing the raw verifier log with the \sys diagnostic raises repair for every model, as shown in Figure~\ref{fig:bpfix-bench}.
On Qwen3.6 27B, one-shot repair rises from 22 of 75 cases with the raw log to 38 with the diagnostic, and one retry raises the two counts to 30 and 44.
GLM 5.2 shows the same pattern, from 28 to 38 in one shot and 52 against 47 with the retry.
The small Qwen2.5 3B repairs none of the 75 from the raw log, and 8 from the diagnostic in one shot and 10 with one retry.
The gain holds across all three models, which points to the diagnostic as its source.
 
To locate the gain, we charge each failed one-shot candidate to the first stage it fails: returning a program, compiling, loading through the verifier, passing the functional test, or passing the source-semantics test.
As Table~\ref{tab:failure-stage} shows, the diagnostic mainly reduces verifier-load and source-semantics failures.
On Qwen3.6 27B, load failures fall from 19 to 10 and source-semantics failures from 22 to 16, and GLM 5.2 shows the same two reductions, from 10 to 5 and from 25 to 22, with compile failures staying low for both models.

For the small Qwen2.5 3B, the raw log is often too weak a repair signal: 62 one-shot candidates still fail at verifier load, and three raw-log prompts exceed the model's context window and return no program.
With the shorter \sys diagnostic, model-call failures disappear and verifier-load failures fall to 39, though compile failures rise from 7 to 14.
\sys gives even a small model enough verifier-facing information to attempt more meaningful repairs, while ordinary code-generation errors remain.
\begin{table}[t]
  \centering
  \caption{Where repairs fail on \bench, by stage, for the one-attempt runs. The \sys diagnostic mainly removes verifier-load (Load) and proof failures; compile and model-call (Call) failures stay low. Each entry counts failing cases at a stage, and pass count plus the row total is 75.}
  \label{tab:failure-stage}
  \small
  \setlength{\tabcolsep}{4pt}
  \begin{tabular}{@{}l l r r r r r@{}}
    \toprule
    Model & Input & Compile & Load & Func. & Proof & Call \\
    \midrule
    Qwen3.6 27B & raw  &  3 & 19 &  9 & 22 & 0 \\
                & \sys &  1 & 10 & 10 & 16 & 0 \\
    GLM 5.2     & raw  &  1 & 10 & 11 & 25 & 0 \\
                & \sys &  1 &  5 &  9 & 22 & 0 \\
    Qwen2.5 3B  & raw  &  7 & 62 &  0 &  3 & 3 \\
                & \sys & 14 & 39 &  6 &  8 & 0 \\
    \bottomrule
  \end{tabular}
\end{table}
 
Overall, the quantitative results match the two case studies.
The diagnostic helps most at the stages where a repair must restore the verifier-visible proof, loading through the verifier and preserving the program's source semantics.
These stages are exactly where \sys localizes the lost proof, so the model receives what a repair must restore.
\section{Related Work}\label{sec:related-work}

\noindent\textbf{eBPF verification and safety.}
Prior eBPF tooling decides whether a program is safe or works around the verifier, leaving rejection diagnosis to the developer~\cite{deokar2024empirical}.
Offline re-provers such as PREVAIL establish safety over their own abstract domain~\cite{gershuni2019simple}, verifier audits check the soundness of its operators~\cite{vishwanathan2023verifying, vishwanathan2022sound, sun2024validating}, and other work replaces the verifier with safe Rust~\cite{jia2025rex}, synthesizes passing programs~\cite{zheng2024kgent}, or isolates programs at runtime~\cite{lu2024moat, zhang2024hive, lim2023unleashing, lim2024safebpf}.
None replays the rejection log to reconstruct where the required proof was lost.

\noindent\textbf{Error diagnosis and fault localization.}
Tools that diagnose rejections outside eBPF read facts the checker built: type-error diagnosis reasons over typing constraints~\cite{pavlinovic2014finding, zhang2014toward, seidel2017learning}, spectrum-based fault localization needs many test runs~\cite{wong2016survey, jones2005empirical}, program slicing needs the dependence graph~\cite{weiser1981program}, and unsat-core extraction needs a solver's search artifacts~\cite{cimatti2011computing}.
For eBPF, Pretty Verifier uses regular expressions to map the terminal error to a source line and hint~\cite{rizza2025design}, and BPF Verifier Visualizer provides a frontend that renders per-instruction state interactively~\cite{bpfvv}, but neither localizes where the proof was lost.
\sys reconstructs the proof lifecycle from the verifier's own printed state, recovering register types, scalar ranges, and pointer provenance~\cite{linuxkernel-bpf-verifier}.
\section{Conclusion}
\label{sec:conclusion}

When the verifier rejects an eBPF program, its terminal error marks where verification stopped, which can lie far from where the program lost the required proof.
Reading only the rejection log, \sys reconstructs the required proof's lifecycle from the verifier's per-instruction states and localizes where the proof was lost.
On \bench, the \sys localization raises a downstream model's one-shot repair from 22 to 38 of the 75 cases, with every accepted fix passing the kernel verifier independently of \sys.
\bibliographystyle{ACM-Reference-Format}
\bibliography{reference}

\end{document}